\documentclass[aps,prl,twocolumn,showpacs]{revtex4}
\usepackage{amssymb}
\usepackage{amsmath}
\usepackage{epsfig}

\begin{document}
\title{Delocalization and spreading in a nonlinear Stark ladder}
\author{Dmitry O. Krimer${}^{1}$, Ramaz Khomeriki${}^{1,2}$ and Sergej Flach${}^1$}
\affiliation {${\ }^1$Max-Planck Institute for the Physics of Complex Systems, N\"othnitzer Str. 38, 01187 Dresden, Germany \\
${\ }^2$Physics Department, Tbilisi State University, Chavchavadze 3, 0128 Tbilisi, Georgia} 

\begin{abstract} 
We study the evolution of a wave packet in a nonlinear Schr\"odinger lattice equation subject to a dc bias.  
In the absence of nonlinearity all normal modes are spatially localized giving rise to a Stark ladder with an 
equidistant eigenvalue spectrum and Bloch oscillations. Nonlinearity induces frequency shifts and mode-mode 
interactions and destroys localization. 
With increasing strength of nonlinearity we observe: (I) localization as a transient, with subsequent subdiffusion 
(weak mode-mode interactions); (II) immediate subdiffusion (strong mode-mode interactions);  (III) single site 
trapping as a transient, with subsequent explosive spreading, followed by subdiffusion.   For single 
mode excitations and weak nonlinearities stability  intervals are predicted and observed upon variation of the 
dc bias strength, which affect the short and long time dynamics. 
\end{abstract}
\maketitle

The textbook solid state problem of an electron in a periodic potential with an additional electric field 
(see e.g. \cite{electron}) lead to investigations of Bloch oscillations \cite{BO} and Landau-Zener tunneling \cite{landau} in different physical systems. Optical waves in photonic lattices \cite{OL_lin,OL_nonlin} and ultracold atoms in optical lattices \cite{BEC_lin,BEC_nonlin} are recent examples. If the well depth of the periodic potential is large enough, Landau-Zener tunneling is suppressed, the problem is discretized (Stark ladder) and the resulting eigenvalue problem is explicitly solved in terms of the localized eigenmodes \cite{SL} of the system.  Additional nonlinear terms in the wave equations are generated e.g. by a nonlinear response of the medium (light) or by a mean-field treatment of quantum many-body interactions (BEC). These terms destroy the integrability of the wave equations leading to decoherence and deterioration of Bloch oscillations, which was observed on time scales of a few oscillation periods \cite{BEC_nonlin,OL_nonlin}. In the present work we study and explain the delocalization process induced by nonlinear terms. It leads to an incoherent subdiffusive spreading of any initially localized wave packet.

We consider a discrete nonlinear Schr\"odinger equation (DNLS) with a dc bias $E$:
\begin{equation}
i\dot\Psi_n=-\left(\Psi_{n+1}+\Psi_{n-1}\right)+nE\Psi_n+\beta|\Psi_n|^2\Psi_n,
\label{eq1}
\end{equation}
where the dot denotes the time derivative, $\Psi_n$ is a complex amplitude of a wave at lattice site $n$, and $\beta$ is the 
nonlinearity strength. Varying $\beta$ is strictly equivalent to varying the norm $\mathcal{N}=\sum_n|\Psi_n|^2$. We choose 
$\mathcal{N}=1$. 

\begin{figure}
{\epsfig{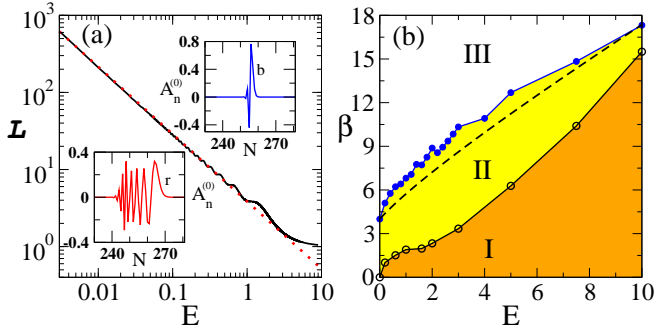}} 
\caption{(Color online) (a) Black solid line: localization volume $\mathcal{L}$ of the eigenfunction $A_n^{(0)}$ versus $E$.  
Red dotted line: The fit of $\mathcal{L}$ by $a\cdot E^{-b}$ with $a=3.847$ and $b=0.874$. Insets: explicit form of 
the eigenfunction across the chain for two values $E=2$ and  $E=0.2$ [(b), blue; (r), red]; (b) The diagram of the 
three regimes of spreading in the parameter space $(\beta,E)$. 
Empty and filled circles: 
numerically obtained thresholds which separate the three different regimes I,II,III.
Lines connecting symbols guide the eye. 
Black dashed line: threshold between II and III  obtained from the dimer model [see text]. In the limit of $E\rightarrow\infty$ 
all lines merge to the asymptotic limit $\beta \propto E$.}
\vspace*{-4mm}
\label{fig1}
\end{figure}

For $\beta=0$ Eq.~(\ref{eq1}) is reduced to an eigenvalue problem using the ansatz $\Psi_n(t)=\exp(-\lambda t)A_n$. In the 
case of an infinite lattice the eigenvalues $\lambda_\nu=E\nu$ (with $\nu$ being an integer) form an equidistant spectrum 
which extends over the whole real axis - the Stark ladder. The corresponding eigenfunctions or normal modes (NMs) obey the 
generalized translational invariance $A_{n+\mu}^{(\nu+\mu)}=A_{n}^{(\nu)}$ 
\cite{BO}. With $J_{n+1}(x)+J_{n-1}(x)=(2n/x)J_n(x)$ for the Bessel function $J_n(x)$ of the 
first kind \cite{book} it follows \cite{SL}
\begin{eqnarray}
A_n^{(0)}=\left\{\begin{array}{l}
(-1)^{n}J_{-n}(2/E)\ \ ,\ \ n\leq 0\ ;\\
J_{n}(2/E)\ \ ,\ \ n\geq 0\ .
\end{array}\right.
\label{eq2}
\end{eqnarray}
All NMs are spatially localized with an asymptotic decay $|A_{n \rightarrow \infty}^{(0)}| \rightarrow \left(1/E\right)^n/n!$, 
giving rise to the well-known Bloch oscillations with period $T_B = 2\pi / E$. In Fig.~\ref{fig1}a  we plot the localization 
volume $\mathcal{L}=1/\sum_n |A_n^{(0)}|^4$ of an eigenstate, which characterizes the spatial extent of 
the NM, as a function of $E$ . In the wide region of $E\in [0.003,1]$ we find
$ \mathcal{L} = a\cdot E^{-b}\;,\;a\approx 3.847\;,\;b\approx 0.874$.  Note that for $E=10$, $\mathcal{L}$ is already close to 
its asymptotic value $\mathcal{L}=1$ for $E\rightarrow \infty$.
The eigenvalue spacing $E$ and the eigenvalue variation over a localization volume $\Delta \equiv E\mathcal{L}$ determine 
the packet evolution in the presence of nonlinearity.

With
\begin{equation}
\Psi_n(t)=\sum\limits_{\nu}\phi_\nu(t)A_n^{(\nu)}e^{-i\nu E t},
\label{eq4}
\end{equation}
Eq.~(\ref{eq1}) transforms into NM space
\begin{equation}
i\dot\phi_{\nu}=\beta\!\!\sum\limits_{\nu_1,\nu_2,\nu_3}I_{\nu,\nu_1,\nu_2,\nu_3}\phi_{\nu_1}^*\phi_{\nu_2} \phi_{\nu_3}e^{i(\nu+\nu_1-\nu_2-\nu_3)\,Et}
\label{eq5}
\end{equation}
where
\begin{eqnarray}
I_{\nu,\nu_1,\nu_2,\nu_3}\equiv\sum\limits_{n} A_{n-\nu}^{(0)} A_{n-\nu_1}^{(0)} A_{n-\nu_2}^{(0)} A_{n-\nu_3}^{(0)}
\label{eq_overl}
\end{eqnarray}
are the overlap integrals between the eigenfunctions (\ref{eq2}).  To characterize the wave packet dynamics we compute: 
i) the participation number $P=1 / \sum_{\nu} |\phi_{\nu}|^4$ which measures the number of strongest excited NMs; 
ii) the second moment $m_2=\sum_\nu (\nu-\bar\nu)^2|\phi_\nu|^2$ ($\bar\nu=\sum_\nu \nu|\phi_\nu|^2$) which quantifies 
the degree of spreading of the wave packet and iii) the compactness index $\zeta=P^2/m_2$ which 
measures the degree of sparcity of the wavepacket. We use the SBAB$_2$ symplectic integrator \cite{LR01} 
to integrate Eq.~(\ref{eq1}) numerically (see also \cite{flach} for details). The number of sites is chosen such 
as to exclude finite size effects in the evolution of the wave packets (typically from $N=500$ to $N=1000$). 

First, we study a single site initial excitation $\Psi_n(0)=\delta_{n0}$. In that case the amplitudes in NM space are  
$\phi_\nu(0)=J_\nu(2/E)$. The nonlinear frequency shift at site $n=0$ is $\delta \equiv \beta$. We expect three 
qualitatively different regimes of spreading: (I) $\delta < E$; (II) $E < \delta < \Delta$; (III) $ \Delta < \delta$. 
In case (I) the nonlinear frequency shift is less than the spacing between excited modes. Therefore no initial resonance 
overlap is expected, and the dynamics may evolve as the one for $\beta=0$ at least for long times. In case (II) 
resonance overlap happens, and the packet should evolve differently. For case (III) $\delta$ tunes the excited site out 
of resonance with the neighbouring NMs. Resonances with more distant NMs are possible, but the overlap with these NMs 
is the weaker the further away they are (\ref{eq2}). Therefore for long times the excited site may evolve as an 
independent oscillator (trapping). The diagram of the three regimes of spreading in the parameter space $(\beta,E)$ is 
shown in Fig.~\ref{fig1}b.
\begin{figure}
{\epsfig{file=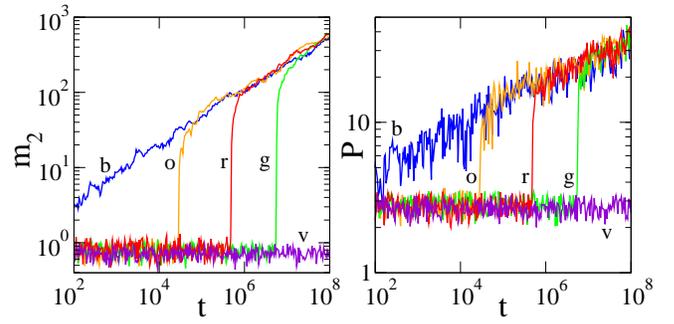,width=1.\linewidth}} 
\caption{(color online) Single site excitation for $E=2$. Second moment $m_2$ and participation number $P$ versus time 
in log-log plots for different values of $\beta$ inside the interval where an explosive delocalisation of the trapped 
regime occurs: $\beta=8.15,8.25,8.5$  [(o) orange; (g) green; (r) red]. $\beta=8$ [(b), blue]: intermediate regime. 
$\beta=8.9$ [(v), violet]: trapped regime.}
\label{fig2}
\end{figure}
\begin{figure}
{\epsfig{file=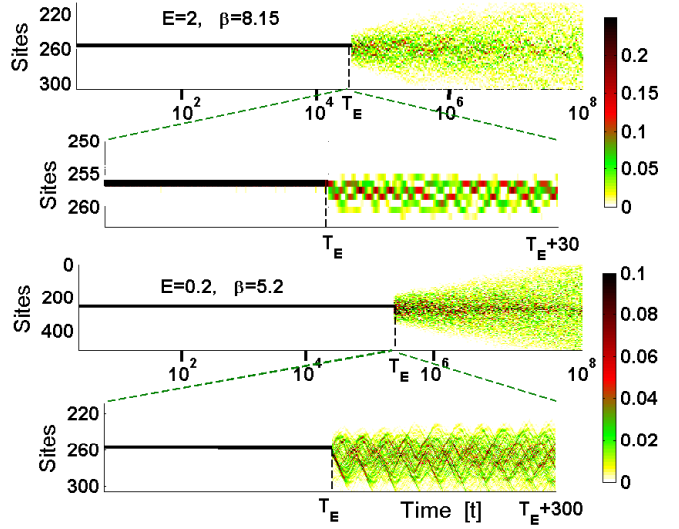,width=1.\linewidth}} 
\caption{$|\Psi_n(t)|^2$ as a function of time (in logarithmic scale) and lattice site for 
$E=2,\, \beta=8.15$ (upper graph) and $E=0.2,\, \beta=5.2$ (lower graph).  
The zooms into the explosion region are plotted on a linear time scale. Bloch oscillations are observed right after
the explosion time $T_E$.}
\label{fig3}
\vspace*{-3mm}
\end{figure}

Let us start with case (III). For $E=2$ and $\beta>8.9$, the single site excitation stays trapped up to times $t=10^8$  
without significantly spreading into any other site of the lattice (violet curve in Fig.\ref{fig2}). Slightly lowering 
$\beta$ we observe that the excitation is trapped up to a some time $T_E$ which sensitively depends on $\beta$ and changes 
by many orders of magnitude e.g. between $10^2$ to $10^7$ in the narrow interval $\beta\in (8.05,8.9)$ for $E=2$ 
(Fig.\ref{fig2}). For times $t > T_E$ an explosive and spatially asymmetric spreading is observed on a time scale of one 
Bloch period $T_B$. The packet spreads in the direction of NMs with larger eigenvalues, which provide the possibility of 
resonant energy transfer from the single site excitation due to its positive nonlinear frequency shift $\delta$. For about 
ten Bloch periods $T_B$ the packet shows  Bloch oscillations, which then quickly decohere. Finally the packet spreads 
incoherently and subdiffusively (Figs.\ref{fig2},\ref{fig3}). The explosion time $T_E$ is not monotonously changing with 
$\beta$ (cf. Fig.\ref{fig2}), which indicates intermittency, i.e. the single site excitation can be closer or farer away from 
some regular structures in phase space. That distance may in turn control the value of $T_E$. 
For $E=2$ and $\beta=8$ the packet spreads from scratch (blue curve in Fig.\ref{fig2}). 
\begin{figure}
{\epsfig{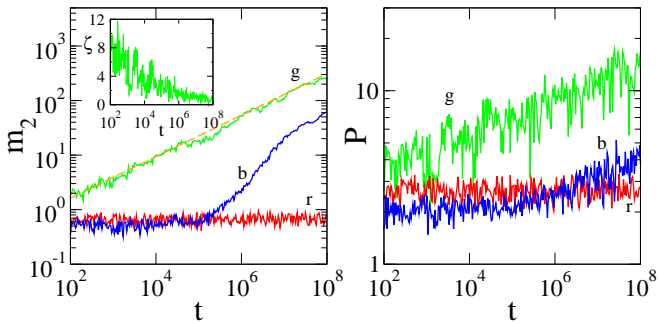}} 
\caption{(color online) Single site excitation for $E=2$.  $m_2$ and  $P$ versus time (log-log plots) for i) $\beta=2.1$ 
[(b), blue]: weak nonlinear regime;  ii) $\beta=4$  [(g), green]: intermediate regime and iii) $\beta=9$ [(r), red]: 
trapped regime. Orange dashed line: $m_2\sim t^{0.38}$. Inset: $\zeta$ for $\beta=4$. }
\label{fig4}
\vspace{-3mm}
\end{figure}

In regime II the nonlinearity is not strong enough to trap a single site excitation, but the excited NMs are resonantly 
interacting. Therefore the packet expands subdiffusively from the very beginning (green curve in Fig.~\ref{fig4}). 
Since the eigenmodes of the linear system are spatially localized, the incoherent way of subdiffusive spreading resembles 
the one observed in disordered nonlinear systems \cite{flach,she}.  The second moment $m_2$ appears to grow without bounds 
according to $m_2 \sim t^{\alpha}$. We fitted the exponent $\alpha$ for 20 different runs for $E=2$ and $\beta=4$. Each 
run was obtained as follows. We integrated the equations up to $t=10^2$.  The obtained  
distribution was transformed to NM space.  The phases of all NMs were randomized, but the amplitudes were kept constant. 
Finally, we performed an inverse transformation to real space and continued the integration up to $10^8$. As a result we 
obtained the average exponent $\alpha=0.38\pm 0.015$. Let us reconsider (\ref{eq5}) and keep only resonant terms 
$\nu+\nu_1-\nu_2-\nu_3=0$. The resulting resonant normal form equations read
\begin{equation}
i\dot\phi_{\nu}=\beta\sum\limits_{\nu_1,\nu_2}I_{\nu,\nu_1,\nu_2,(\nu+\nu_1-\nu_2)}\phi_{\nu_1}^*\phi_{\nu_2} 
\phi_{\nu+\nu_1-\nu_2}\;.
\label{eq5b}
\end{equation}
If at least two neighbouring NMs are excited, the resonant normal form (\ref{eq5b})
will have a connectivity similar to the original lattice equations, which allows spreading over the whole lattice. 
Excluding the possibility of integrability of (\ref{eq5b}), we expect this spreading to happen indeed, as observed 
in the numerical runs. Following \cite{flach} we conclude, that all NMs are in resonance with their neigbourhood. 
If the wave packet is in a state of equipartition at any time, it follows $\alpha=0.5$ \cite{flach}. In such an 
assumed equipartitioned state the compactness index $\zeta \approx 3$ \cite{flach}. However, the compactness index 
$\zeta$ decreases in time way beyond the equipartition threshold (inset in Fig.~\ref{fig4}) and therefore the 
assumption of equipartitioned wave packets is incorrect. This is the reason for the discrepancy between the expected 
(0.5) and observed (0.38) values of the exponent $\alpha$.

In regime I the nonlinear frequency shift $\delta< E$. No strong initial resonance overlap between NMs is expected. 
In that case the dynamics may (at least for long times) 
evolve as in the linear case.  Eventually, the subdiffusive spreading takes place after some waiting time $\tau$ (Fig.~\ref{fig4}).  
Note that the crossover from localization to spreading around time $\tau$  is not accompanied by a rapid redistribution of 
the energy between many neighboring sites as in the case of trapping with large nonlinearities. Therefore the onset of 
subdiffusion is simply related to the fact, that weakly chaotic dynamics is characterized by a small momentary diffusion rate, 
and the corresponding diffusion process is observed only after times larger than $\tau$.  

In Fig.~\ref{fig1}b we plot the boundaries between the three regimes in the parameter space ($\beta,E$). These boundaries 
are determined as follows. Keeping the value of $E$ fixed and varying $\beta$, we integrate the equations up to $t=10^5$ 
starting from the single site excitation. Then we calculate $m_2$ and $P$ of the final wave packet. Being in regime I, these 
values are quite close to those for the linear case.  For a critical value of $\beta$ they 
start to grow indicating that we enter regime II.  Finally, at a second and larger
value of $\beta$ the values of
$m_2$ and $P$ drop down again, signalling the entering of regime III.

The border between regime II and III can be accounted for by using a dimer model.  
Since the trapped state corresponds to only a single site being excited, 
we consider a dimer model where only one further lattice site (to the right)
is added, in order to model the asymmetric energy transfer during the explosion. The resulting equations read
\begin{equation}
i\dot\Psi_0=-\Psi_1+\beta|\Psi_0|^2\Psi_0,\, 
i\dot\Psi_1=-\Psi_0+E\Psi_1+\beta|\Psi_1|^2\Psi_1,
\nonumber
\end{equation}
which are solved using elliptic functions \cite{smerzi}. 
The dimer model is integrable, and the analytical computation of the initial condition 
$|\Psi_n(0)|^2=\delta_{n,0}$
crossing a separatrix
in the phase space of the dimer yields a relation between $\beta$ and $E$.
The transition to the trapped state at some critical nonlinearity 
$\beta(E)$ is accompanied by divergence of the period of oscillations and above the threshold the energy is mostly 
concentrated on one site. The final equation to be solved numerically reads
\begin{equation}
-\beta^3E+\beta^2(3E^2-1)+\beta E(20-3E^2)+(E^2+4)^2=0.
\label{d3}
\end{equation}
As is seen from Fig.\ref{fig1}b it fits very well with the boundary between regime II and III
obtained from the evaluation of the full set of equations.
In particular, (\ref{7}) correctly predicts that $\beta(E\rightarrow 0) \rightarrow 4$
and $\beta(E \rightarrow \infty) \rightarrow E$.

\begin{figure}
{\epsfig{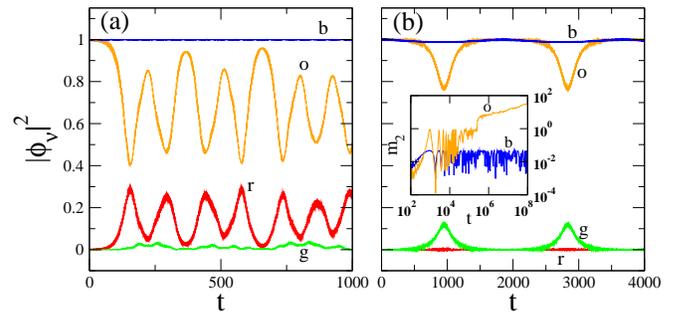}} 
\caption{(color online) Amplitudes of the excited,  $\nu=0$, and neighboring modes, $\nu=1$ and $\nu=2$,  versus time for $\beta=0.4$. 
(a)  E=1.6, $\nu=0$ [(b), blue], E=2, $\nu=0,1,2$ [(o), orange; (r), red; (g), green]. (b)  E=1.18, $\nu=0$ [(b), blue]; 
E=1.26, $\nu=0,1,2$ [(o), orange; (r), red; (g), green]. Inset: long time evolution. $m_2$ versus time in log-log plots: 
E=1.18,1.26  [(b), blue; (o), orange]. } \label{fig5}
\vspace*{3mm}
\end{figure}

A single mode excitation $\phi_\nu(t=0)=\delta_{\nu,0}$ also exhibits the three different regimes of spreading. However, for 
small values of nonlinearity $\beta$ a new intriguing feature of the short time dynamics follows. Indeed, consider the 
resonant normal form (\ref{eq5b}).
For a single mode excitation, it follows strictly that no other NM is going to be excited.
Therefore it is an exact solution of (\ref{eq5b}) 
\begin{equation}
\phi_0(t)=\exp\left(-i\beta I_{0,0,0,0}t\right).
\label{eq8}
\end{equation}
However, the full set of equations (\ref{eq5}) will excite other NMs as well. These small perturbations may stay small or 
start to grow, depending on the stability of the solution (\ref{eq8}) of equations (\ref{eq5b}). The linear equations for 
small perturbations with $\nu \neq 0$ read
\begin{eqnarray} \label{7}
i\dot\phi_{\nu}&=&\beta\left[I_{-\nu,\nu,0,0}\phi_0^2 \phi_{-\nu}^* +2I_{\nu,\nu,0,0} \phi_{\nu}\right], \\
i\dot\phi_{-\nu}^*&=&-\beta\left[I_{-\nu,\nu,0,0}\phi_0^{*2} \phi_{\nu} +2I_{\nu,\nu,0,0} \phi_{-\nu}^*\right]. \nonumber
\end{eqnarray}
Note that only pairs of NMs with $\pm \nu$ are coupled. With $\phi_{\pm \nu}\sim \exp\left(-i\beta I_{0,0,0,0}t+\gamma_\nu t\right)$ 
one gets
\begin{equation}
\gamma_{\nu}=|\beta|\sqrt{I_{-\nu,\nu,0,0}^2-\left(2I_{\nu,\nu,0,0}-I_{0,0,0,0}\right)^2}.
\label{eq12}
\end{equation}
Thus if $I_{-\nu,\nu,0,0}^2 > \left(2I_{\nu,\nu,0,0}-I_{0,0,0,0}\right)^2$ it follows that  $\gamma_\nu$ is real, corresponding 
perturbations will grow exponentially fast in time, and the original single mode excitation is unstable. Since the overlap 
integrals depend on the dc bias $E$, we expect transitions from stability to instability upon varying $E$. Indeed, a numerical 
evaluation of that condition yields that the single mode excitation is stable in the limits: $1.134<E<1.23$,  $1.314<E<1.917$, $E>2.314$; 
otherwise it is unstable. In the instability window $1.917<E<2.314$ ($1.23<E<1.314$) the modes $\nu=\pm 1$ ($\nu=\pm 2$) 
become unstable whereas the other modes remain stable.  This behavior is confirmed by numerical simulations presented in 
Fig.~\ref{fig5} on short time scales. After sufficient exponential growth the amplitude of the unstable mode saturates due 
to nonlinear corrections. Due to recurrences 
the energy eventually comes back to the  initially excited mode after possible long times.
We also find a correlation between the short time stability properties of a single mode excitation and the long time evolution of 
the wave packet. If the single mode excitation is launched in a stability window, the wave packet is practically not spreading 
up to $t = 10^8$
(inset in Fig.\ref{fig5}). However, a small change of the dc bias value $E$ can tune the system into an instability window. 
As a result, the wave packet shows subdiffusive spreading on the accessible time scales,
starting with $t \sim 10^5$ (inset in Fig.\ref{fig5}). 
In conclusion, we have investigated various regimes of wave packet spreading for the nonlinear Stark ladder. 
For large nonlinearities there is a long lived trapped regime with an explosive 
transition to Bloch oscillations, followed by a subdiffusive spreading at large time scales. For intermediate 
nonlinearities we find subdiffusion from scratch.  For small nonlinearities the system is showing localization as a transient, 
with a subsequent subdiffusive spreading.  For single mode excitations additional stability and instability intervals with 
respect to the dc bias strength exist. For weak nonlinearity this leads to an additional sensitive dependence of the 
transient localization times on the value of the dc bias. In all cases, we observe that nonlinearity destroys integrability, 
introduces chaos, and ultimately leads to a destruction of localization, which was a hallmark of the linear system due to phase 
coherence and destructive/constructive wave interference.
 
\paragraph{Acknowledgements.} The authors thank A. Ponno and Ch. Skokos for helpful discussions and comments. R. Kh. acknowledges 
financial support of the Georgian National Science Foundation (Grant�No GNSF/STO7/4-197).

\end{document}